\documentclass[preprint,showpacs,preprintnumbers,aps,prb,floatfix]{revtex4}
\usepackage{graphicx}

\begin{document}
%
%
\title{Intensity of Brillouin light scattering from spin waves in
magnetic multilayers with noncollinear spin configurations: Theory
and experiment}

\author{M. Buchmeier,\cite{mbuch} H. Dassow, D. E. B\"urgler,
and C. M. Schneider}
\affiliation{Institut f\"ur Festk\"orperforschung and cni - Center of
Nanoelectronic Systems for Information Technology,
Forschungszentrum J\"ulich GmbH, D-52425 J\"ulich, Germany}

\begin{abstract}
%
%
The scattering of photons from spin waves (Brillouin light scattering -- BLS) is a well-established technique for the study of layered magnetic systems. The information about the magnetic state and properties of the sample is contained in the frequency position, width, and intensity of the BLS peaks. Previously [Phys.\ Rev.\  B {\bf 67}, 184404 (2003)], we have shown that spin wave frequencies can be conveniently calculated within the ultrathin film approach, treating the intralayer exchange as an effective bilinear interlayer coupling between thin virtual sheets of
the ferromagnetic layers.  Here we give the consequent extension of this approach to the calculation of the Brillouin light scattering (BLS) peak intensities. Given the very close relation of the BLS cross-section to the magneto-optic Kerr effect (MOKE), the depth-resolved
longitudinal and polar MOKE coefficients calculated numerically via the usual magneto-optic formalism can be employed in combination with the
spin wave precessional amplitudes to calculate full BLS spectra for a given magnetic system. This approach allows an easy calculation of BLS intensities even for noncollinear spin configurations including the exchange modes. The formalism is applied to a Fe/Cr/Fe/Ag/Fe trilayer system with one antiferromagnetically coupling spacer (Cr). Good agreement with the experimental spectra is found for a wide variety of spin configurations.
\end{abstract}

\date{\today}
\maketitle
%
%
\section*{Introduction}
\label{intro}
In a Brillouin light scattering (BLS) experiment spin waves in a magnetic system are
probed via inelastic scattering of photons.  The spin wave mode or magnon
appears as a peak in the recorded spectrum, which is shifted by the
magnon frequency $\nu_m$ relative to the central peak in the spectrum, which in turn is caused by
elastically scattered photons.  The shift reflects either an energy loss or energy gain
corresponding to the creation (Stokes condition) or annihilation
(anti-Stokes condition) of a magnon, respectively. The spectrum
contains three different types of information: (i) the peak positions,
which are determined by the spin wave frequencies, (ii) the peak shapes and
especially their widths, which are related to the lifetime of the
magnons, and (iii) the peak areas, which are proportional to the
scattering cross-sections.

Most experiments focus on an analysis of the spin wave frequencies, which contain information about many
magnetic properties such as, for instance, saturation magnetization,
anisotropies, and interlayer coupling.  With a suitable
experimental geometry and procedure, these properties can be determined solely on the basis of the magnon frequencies. The peak width or linewidth contains information about the
spin wave lifetime $\tau$ as it results in a linewidth broadening
proportional to $1/\tau$.  However, in most cases concerning epitaxial
samples this magnon linewidth is much smaller than the apparatus broadening
of about 1\,GHz and cannot be resolved.  Only when the magnon lifetime
is strongly reduced, \textit{e.g.} due to extrinsic two-magnon
scattering in the case of exchange bias,\cite{PRB_63_214418}
interlayer exchange coupling,\cite{JAP_93_3427} or in polycrystalline
samples,
the magnon linewidth becomes larger than the
apparatus resolution and can be analyzed.  On the other hand, the
\emph{scattering intensities}, which are the topic of this work, carry
information mainly about the precessional amplitudes of the spin waves,
but also depend on the optic and magneto-optic properties of the
sample. For a full quantitative analysis of the BLS spectra, however, peak positions, line widths, and peak intensities have to be treated on an equal footing.

It is important to note that very few publications
\cite{JPC_8_211,JAP_50_7784,PRB_18_4821,PRB_22_5420,JMMM_73_299,JAP_64_6092,JAP_69_5721,
PRB_42_508,PRB_64_134406,PSS_196_16,PRB_53_2627,SS_454_880,SS_507_502,
PRB_65_165406,PRB_63_104405,JMMM_290_530} have yet addressed the issue of the
scattering intensities, although they contain valuable information
about the mode types, the alignment of the magnetic moments and can
even be used to investigate the magneto-optic
coupling.\cite{PRB_63_104405} This lack of interest is
probably due to the fact that the cross-sections cannot be easily
analyzed in an intuitive manner, but require a comparison with theory.
On the other hand, the kind of detailed information hidden in the peak intensities is of high
relevance for many technological applications, such as data storage
and communication technology, because the operation frequencies of contemporary
magnetic devices approach the GHz regime, where the magnetization
dynamics is closely related to the spin wave modes.

The computation of the scattering
intensities is commonly considered to be very
complicated.\cite{JMMM_73_299}
There is only a handful of research groups, which have successfully
developed and applied a suitable formalism, although the computation scheme is
now known for more the three decades.\cite{JPC_8_211}
Part of the computational complexity stems from the so-called partial waves
approach,\cite{JMMM_73_299,JMMM_82_186,PRB_41_530,JMMM_145_278} which
has been used to calculate the spin wave cross-sections in most
publications.  To our knowledge this approach has not yet been applied to
and solved for noncollinear alignments of the magnetization and, thus,
studies dealing with BLS intensities in noncollinear states are rarely
available.\cite{PRB_42_508,PRB_53_2627,JMMM_290_530} 
However exactly those noncollinear states
are crucial for the magnetization switching
employed in a large number devices, such as for instance MRAM. 
A detailed understanding of the dynamics of noncollinear states 
is therefore highly desirable and a theoretical 
description of the spinwave properties of noncollinear states 
is of high technological relevance.

Only three publications
\cite{PRB_42_508,PRB_64_134406,PRB_53_2627} employ the much easier
ultrathin film approximation
(UTFA),\cite{PRB_42_508,PRB_49_339,PRB_54_3385,JAP_84_958} which
has been thought 
to be accurate only for the lowest-energy
modes of ferromagnetic films with thicknesses well below the exchange
length, \textit{i.e.} a few nanometers in the case of Fe.
However, in our previous work (Ref.\ \onlinecite{PRB_67_184404}) we have shown that the
UTFA approach can be easily extended in order to accurately describe
the modes in thicker films. We thereby included also the exchange modes and the
twisted ground state found in exchange springs and also in the case of
very strong interlayer exchange coupling.  In the present contribution, we elucidate how this
calculation scheme can be further extended to describe BLS
cross-sections.  In fact, the cross-section calculation using this
method only requires a standard treatment of the
light propagation in the multilayer apart from the formulae already presented in Ref.\
\onlinecite{PRB_67_184404} .  As we will show below, the
optics involved is closely related to the magneto-optic Kerr effect
(MOKE). This has the advantage that
existing computer codes for the calculation of the MOKE
can be reused to evaluate the BLS intensities.

Section \ref{calc} gives a detailed description of our computational
procedure.  In order to demonstrate the accuracy of the scheme we have
recomputed the spectrum of Ref.\ \onlinecite{JMMM_73_299} and find an
excellent agreement with respect to the peak positions and intensities.  An open-source computer program for the
calculation of BLS frequencies, precessional amplitudes, and
intensities (although excluding the ground state calculation) as well
as for the calculation of MOKE can be downloaded from our
server, see Ref.\ \onlinecite{code}.

In Sec.\ \ref{section2} we apply our formalism to an well
characterized epitaxial spin valve structure of the sequence
Fe(14nm)/Cr(0.9\,nm)/Fe(10\,nm)/Ag(6\,nm)/Fe(2\,nm).  
While the bottom Fe/Cr/Fe trilayer couples
antiferromagnetically, the top, thin Fe layer is decoupled and can be
switched more easily by an external field.  As a consequence this
layer stack shows a rich variety of different spin configurations as a
function of field strength and orientation.  
The application of BLS intensity theory to such a model system
with multiple well defined spin configurations has to
our knowledge been lacking yet.
We find excellent
agreement between our model calculation and the experimental spectra
for all spin configurations.
\section{Calculation}
\label{calc}
\begin{figure}[b]
\centering\leavevmode
\includegraphics[width=0.9\linewidth,clip]{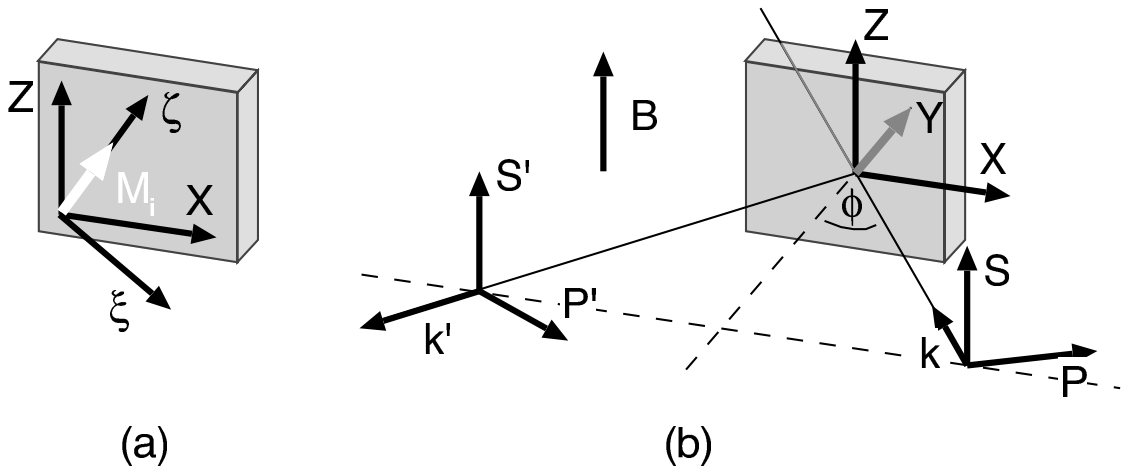}
\caption{(a) Relation between the laboratory frame ($xyz$) and the
coordinate system (${\xi}y{\zeta}$) attached to the static
magnetization $M_{i}$ in each sheet.  (b) Scattering geometry.}
\label{coord}
\end{figure}
%
%
The calculation of the spin wave frequencies and scattering intensities
is carried out based on our approach described in detail in Ref.\
\onlinecite{PRB_67_184404} and requires only little additional
programming effort.  Only the standard formalism for the calculation of
the MOKE in multilayers either using the full $4\times4$
\cite{Yeh,CJP_41_663,PRB_43_6423} or the approximate $2\times2$
\cite{PRB_64_235421} matrix approach is needed in addition to the
formalism in Ref.\ \onlinecite{PRB_67_184404}.  The computation is
carried out in a sequence of four steps as explained in the following:

\subsection{Calculation of the magnon frequencies}
The calculation of the magnetic ground state and the spin wave
frequencies in the extended UTFA is
described in detail in Ref.\ \onlinecite{PRB_67_184404}.
In brief, the intralayer exchange within the ferromagnetic layers is
treated as an effective bilinear interlayer coupling $J^{eff}_{1}$
between thin virtual sheets of the ferromagnetic material.  The
virtual multilayer consisting of the virtual ferromagnetic sheets
is then treated within the conventional UTFA
approach.\cite{PRB_42_508,PRB_49_339,PRB_54_3385,JAP_84_958} 
It is of crucial importance to divide the
ferromagnetic layers into thin enough sheets in order to obtain (i) the
correct ground state for the case of strong interlayer coupling or (ii)
the frequencies of modes, which are not uniform in the perpendicular
direction, \textit{i.e.} Damon-Eshbach modes, optic and exchange
modes.  In particular, an improper ground state configuration can lead
to the disappearance of some modes, especially the optic modes in
antiferromagnetically coupled systems.

%
\subsection{Calculation of the precessional amplitudes}
\label{calc_descr}
The calculation of the mode profiles is straightforward.  For a
multilayer with $N$ sheets, the numerical procedure consists of
solving the set of $2N$ equations given by Eq.\ (A2) in Ref.\
\onlinecite{PRB_67_184404} for the dynamic magnetization components
$m_{i,k}$ of each mode with index $n$ and corresponding frequency
$\omega _n$:
\begin{equation}
\sum _{i,k} A_{j,l}^{i,k}(\omega _n) m_{i,k} = 0,
\label{LSE}
\end{equation}
where $i,l$ are sheet indices ($i,l=1\ldots N$) and $k,l$ are axis
indices ($k,l=y,\xi$).  The relation between the laboratory frame
($xyz$) and the coordinate system (${\xi}y{\zeta}$) attached to the
static magnetization $M_{i}$ in each sheet is shown in Fig.\
\ref{coord}(a).  This calculation can be easily carried out with
standard matrix algorithms, \textit{e.g.} from Ref.\ \onlinecite{nr}.

%
\subsection{Normalization of the precessional amplitudes}
In order to calculate the relative scattering intensities the
amplitude of each spin wave mode has to be normalized such that its
energy per unit area is equal to the thermal energy.  For low
frequency modes ($\hbar \omega \ll kT$), which are of relevance for the present case, the thermal energy is just $kT$.
Therefore, the mode amplitudes have to be adjusted so that they have
the same energy.  The mean energy per unit area stored in a mode is
given by
\begin{mathletters}
\label{modeenrgy}
\begin{eqnarray}
E(\vec m_{i}) =
\frac 1 2 \sum _{i,j} \Bigl[
  \frac {\partial ^2 F} {\partial m_{i, y} \partial m_{j, y}} m_{i, y}
m_{j, y} \nonumber\\
+  \frac {\partial ^2 F} {\partial m_{i, \xi} \partial m_{j, \xi}}
m_{i, \xi} m_{j, \xi} \Bigr],
\end{eqnarray}
\end{mathletters}
where $F$ is the magnetic free energy per unit area and $\vec m _i$ is
the dynamic magnetization per unit area inside sheet $i$.  Fortunately
the matrix elements $A_{j,l}^{i,k}$ are proportional to the second
derivatives of the free energy,
\begin{mathletters}
\label{d2F}
\begin{eqnarray}
A_{i, \xi}^{j, y} = \frac{1}{M_{j}d_{j}} \quad
\frac {\partial ^2 F} {\partial m_{i, y} \partial m_{j, y}}  \\
A_{i, y}^{j, \xi} = \frac{1}{M_{j}d_{j}} \quad
\frac {\partial ^2 F} {\partial m_{i, \xi} \partial m_{j, \xi}}.
\end{eqnarray}
\end{mathletters}
The mode energy normalization is thus performed via scaling $\vec m _i$
by $1/ \sqrt{E(\vec m _i)}$.


%
\subsection{Calculation of the scattering intensities}
In the next step the interaction of the spin waves with the electromagnetic wave via the
magneto-optic coupling has to be taken into account.
The spin wave
induces a polarization $\Delta \vec P_{i}$ inside sheet $i$ oscillating
at $\omega _{Light} \pm \omega _{SW}$:\cite{JMMM_73_299}
\begin{equation}
\label{DeltaP}
\Delta \vec P _i =
\Delta \epsilon _i \vec E _i =
\frac 1 2 K _i \vec E _i \times \vec m _i,
\end{equation}
where $\Delta \epsilon _i$ is the magnetic part of the dielectric
tensor, $K _i$ is the linear magneto-optic coupling constant, $\vec E
_i$ is the electric field of the light, and $\vec m _i$ is the dynamic
magnetization, each inside the magnetic sheet with index $i$.  Only
the first order magneto-optic coupling or terms linear in $m$ have
been taken into account.  The phase of the dynamic magnetization has
to be considered, which means $\vec m$ is a complex vector with
\textit{e.g}.  imaginary out-of plane component $m _y$ and real
in-plane component $m _{\xi}$ as assumed in Ref.\
\onlinecite{PRB_67_184404}.\cite{note}
It is worth noting that exactly the same formula (\ref{DeltaP}) can
also be used to calculate the MOKE signal in the approximative $2\times2$
matrix approach of Ref.\ \onlinecite{PRB_64_235421}.  The only
difference between Eq.\ (4.3) in Ref.\ \onlinecite{PRB_64_235421} and
Eq.\ (\ref{DeltaP}) is the factor $1/2$ and the use of the static
instead of the dynamic magnetization.
In both cases of BLS and MOKE, the electromagnetic wave generated by the
induced polarization $\Delta \vec P$ has to be equated taking into
account the dielectric tensors and standard boundary conditions of all
sheets.  The procedure is exactly the same for the reflected and the
$180^{\circ}$ backscattered light.  The only difference is that the
reflected light propagates along the $\pm \vec k'$ directions, while
the $180^{\circ}$ backscattered light propagates along the $\pm \vec
k$ directions [see Fig.\ \ref{coord}(b)] corresponding to a reversed
sign of the in-plane component of the wavevector.

In the case of MOKE it is well known that the resulting complex Kerr
angle, \textit{e.g.} for incident s-polarization $\Phi _s = E' _p / E'
_s = r _{sp} / r _{ss}$ can be expanded as a function of the Cartesian
magnetization components:\cite{JAP_91_7293}
\begin{equation}
\label{phenMOKE}
\Phi _s
= \sum _i \Bigr[ r _{sp, (M=0)} + L _i m _{i, x} + P _i m _{i, y}
+O(m^2) \Bigl] / r _{ss},
\end{equation}
where the diagonal reflection coefficient $r _{ss}$ is independent of
the magnetization, and the nonmagnetic part of the off-diagonal
reflection coefficient $r _{sp,(M=0)}$ is zero for materials with
cubic symmetry.  Terms quadratic in $m$
are known to be small unless the angle of incidence
is nearly perpendicular or grazing.  Terms linear in the transversal
magnetization component $m _z$ only appear, if the incident light has
p-polarized components. 

From the above discussion it is clear that the intensity of the
scattered light from each mode with index $n$ can also be decomposed
in the same way as in Eq.\ (\ref{phenMOKE}):
\begin{equation}
\label{phenBLS}
I ^{(n)} = \left|E ^{(n)} _{BLS} \right|^2
= \frac{I _0}{4} \left|\sum _i ( L ^{BLS} _i m _{x, i} ^{(n)} + P ^{BLS}
_i m _{y, i} ^{(n)}) \right|^2.
\end{equation}
Equations (\ref{phenBLS}) and (\ref{phenMOKE}) can easily be deduced
employing the approximate magneto-optic formalism in Ref.\
\onlinecite{PRB_64_235421}.
Additionally, it can be shown that the coefficients $L$ and $P$ in
Eqs.\ (\ref{phenMOKE}) and (\ref{phenBLS}) in fact only differ by
sign. This is also plausible from simple symmetry considerations:
Neglecting the small Kerr component and assuming s-polarization of the
incident light the electric field vector is parallel to the
$z$-direction everywhere.  Therefore, the polar magnetization component
$m _y$ results in a Kerr component with electric field vector parallel
to $\Delta P \propto E _z \times m _y$, that is parallel to the
$x$-direction.  As the projection of $\hat x$ on the $P$ and $P'$
directions is the same [see Fig.\ \ref{coord}(b)], the polar
coefficient must be the same for reflection and backscattering,
\textit{i.e.} $P _{BLS} = P _{MOKE}$.  On the other hand, the
longitudinal magnetization component $m _x$ results in an electric
field vector of the Kerr component parallel to the $y$-direction.  As
can be seen from Fig.\ \ref{coord} (b) the projections of the $y$-direction
on $P$ and $P'$ only differ in sign.  Therefore, the longitudinal
coefficient has opposite sign for the $180 ^{\circ}$ backscattering
configuration, \textit{i.e.} $L _{BLS} = - L _{MOKE}$.  The
arguments are more involved in the case of incident
p-polarization.  However, it is well-known from theory that the
first-order intensities are the same for sp and ps
scattering.\cite{PRB_64_134406} In summary, the BLS scattering
intensities in the $180 ^{\circ}$ backscattering configuration can be
calculated in the following way:
\begin{equation}
\label{phenBLS2}
I ^{(n)} \propto
  \left |\sum _i ( - L ^{MOKE} _i m _{x, i} ^{(n)} + P ^{MOKE} _i m _{y,
i} ^{(n)}) \right|^2,
\end{equation}
where the dynamic magnetization of each spin wave mode $n$ has to be
computed and normalized as described above, and $L _i$ and $P _i$ are
the effective longitudinal and polar MOKE coefficients of layer $i$ as
defined by Eq.\ (\ref{phenMOKE}).  They can be calculated using the
standard $4\times4$ formalism as explained in Refs.\
\onlinecite{Yeh,CJP_41_663,PRB_43_6423} or even more easily with the
approximative $2\times2$ approach presented in Ref.\
\onlinecite{PRB_64_235421}.

\subsection{Comparison with the theory of Cochran and Dutcher}
\label{calc_sample}
\begin{figure}[b]
\centering\leavevmode
\includegraphics[width=0.9\linewidth,clip]{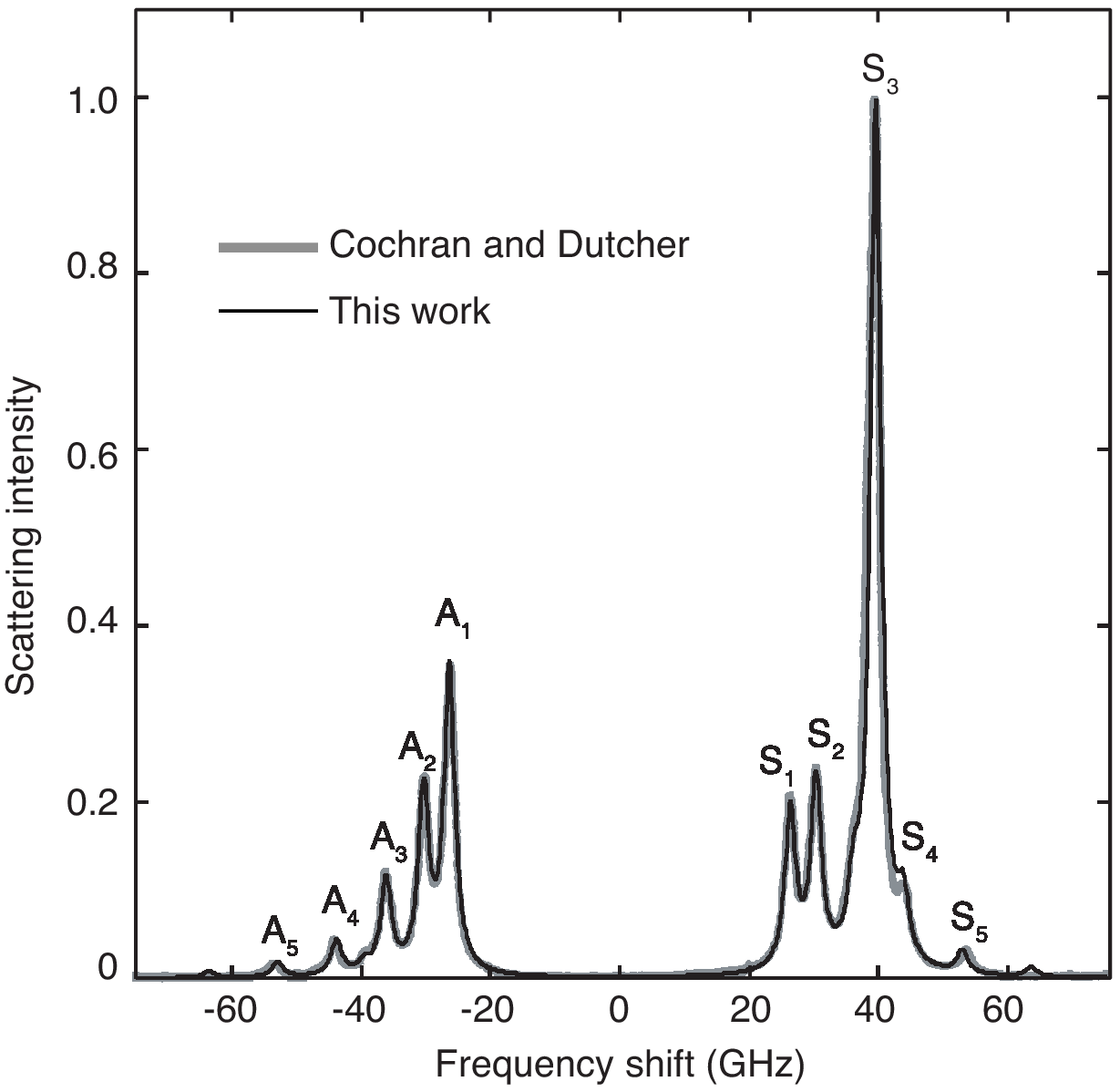}
\caption{Calculated spectrum of a 85\,nm-thick Fe layer (black) in
comparison with the spectrum calculated by Cochran and Dutcher (grey)
reproduced with permission from Fig.\ 2(a) in Ref.\ \protect
\onlinecite{JMMM_73_299}.
 \protect  \footnote{Reprinted from Journal of Magnetism and Magnetic
 Materials, Vol 73,
 J. F. Cochran, and J. R. Dutcher,
 Calculations of the Intensity of Light Scattered from Magnons in Thin
 Films,
 Pages 299 - 310, Copyright (1988), with permission from Elsevier}
}
\label{spectra1}
\end{figure}
In order to check the reliability and accuracy of the calculational scheme described
above, we have calculated the relative intensities of a
thick iron film on a Au substrate using the same parameters as given by Cochran
and Dutcher in Ref.\ \onlinecite{JMMM_73_299}.  The parameters for the
Fe film are: thickness 85\,nm, saturation magnetization $M
_S=1.68$\,MA/m, exchange constant $A = 1.8 \times 10 ^{-11}$\,J/m,
gyromagnetic ratio $\gamma / 2 \pi = 29.26$~GHz/T, and refraction
index $n_{Fe} = 2.83 +2.90i$.  For the Au substrate we use a
refraction index $n_{Au} = 0.67 +2.05i$.  The sample is saturated in
an external field of $0.3$~T applied perpendicular to the plane of
incidence and parallel to the film plane [Fig.\ \ref{coord}(b)].  The
light wavelength of 514.5\,nm and the angle of incidence of
$45^{\circ}$ results in an in-plane magnon vector of $1.73 \times
10^7$~m$^{-1}$.  The absolute value of the magneto-optic coupling or the
presence of a thin capping layer (as usually the case in experimental
studies) are of minor importance as they merely scale the absolute value
of all intensities, while the spectra are anyway normalized to a
maximum intensity of 1 for the anti-Stokes Damon-Eshbach mode labelled
$S _3$.

In order to obtain accurate results the layer has been virtually divided
into 85 sheets each 1\,nm thick.  As in Ref.\ \onlinecite{JMMM_73_299}
the calculated spectrum is convoluted with a Lorentzian lineshape with
a full width at half maximum (FWHM) of 2\,GHz, which corresponds to a
typical instrumental function of a BLS experiment.  The results of our
calculation are shown in Fig.\ \ref{spectra1} and compared with the
spectrum of Cochran and Dutcher.\cite{JMMM_73_299} The agreement is
excellent and the two spectra can hardly be distinguished.  Only the
modes A4, A5, S4 and S5 differ slightly in frequency by approximately 1\,GHz.
This small deviation should not arise from using the extended UTFA,
which only slightly overestimates (less than 0.2\,GHz) the frequency
of the highest modes as compared to the partial waves method.  It
might, however, be due to the fact that we used the magnetostatic
approximation, \textit{i.e.} we did not take into account the electric
high-frequency conductivity in our spin wave calculation. This will cause some error for films with thicknesses comparable or larger
than the skin depth at the spin wave frequency.

\section{Application to a coupled trilayer system}
\label{section2}
In this section, we apply the formalism described in Sec.\ \ref{calc} to
a multilayer structure with three ferromagnetic layers, which are
subject to competing interactions and thus show a rich variety of
spin configurations as a function of the external field.

\subsection{Sample preparation and experimental setup}
\label{prep}
%
Epitaxial Fe(14\,nm)/Cr(0.9\,nm)/Fe(10\,nm)/Ag(6\,nm)/ Fe(2\,nm)
spin valve structures have been prepared by molecular beam epitaxy on
top of a GaAs/Ag(001) buffer system.  The samples have been capped
with a 50\,nm-thick antireflection ZnS layer in order to prevent
oxidation and enhance the magneto-optic response.  This kind of
structures are interesting model systems, which we are also employing to
investigate current-induced magnetization switching.\cite{arxiv} The
preparation is described in detail elsewhere.\cite{arxiv} The Cr
thickness has been chosen in order to obtain a strong antiferromagnetic
coupling in the bottom Fe/Cr/Fe trilayer, which fixes the center
reference layer.  The top, thin Fe layer is decoupled and can be
switched more easily by an external field or a perpendicularly applied
current.  As the samples are fully epitaxial and therefore mainly in a
magnetic single domain state the remagnetization behavior can be
modeled easily.  However, as a consequence of the various competing
interactions -- Zeeman energy, magnetocrystalline anisotropy of all
ferromagnetic layers, interlayer exchange coupling -- a rich variety of
different magnetization configurations is possible and hysteretic
effects determine which configurations actually occur at low fields.

The BLS spectra have been recorded using a Sandercock type ($2\times3$)
pass tandem Fabry-P\'erot interferometer \cite{TAP_51_173} in the
$180^{\circ}$ backscattering geometry.  The wavelength $\lambda =
532$\,nm of the laser light together with the angle of incidence of
45$^{\circ}$ results in an in-plane magnon wave vector $q = 1.67 \times
10^7$~m$^{-1}$ of the measured magnons.  The external field was
applied in the film plane and perpendicular to the magnon wave vector
as sketched in Fig.\ \ref{coord}(b).
\subsection{Results and discussion}
\label{results}
\begin{figure}[hbt]
\centering\leavevmode
\includegraphics[width=0.9\linewidth,clip]{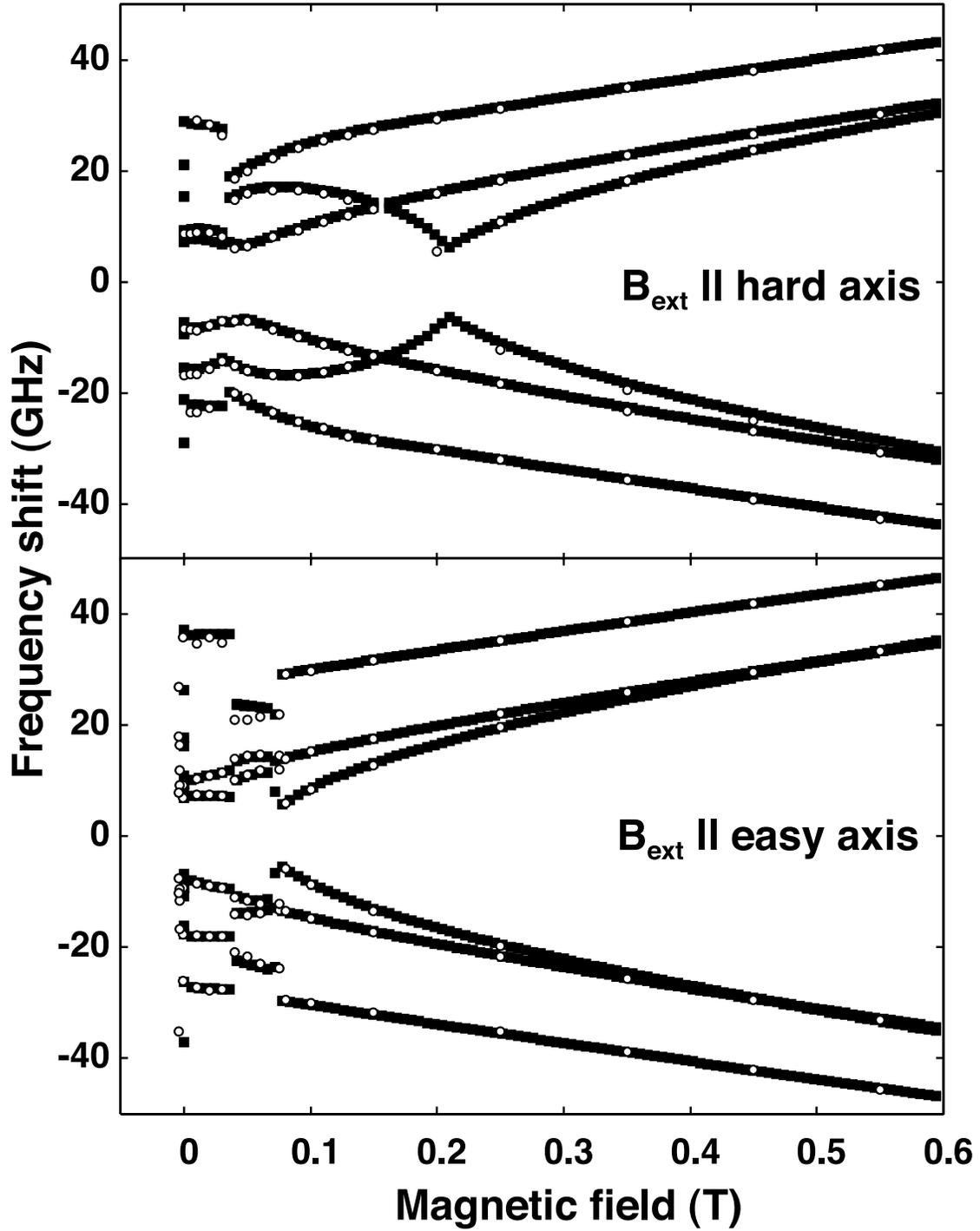}
\caption{Field dependence of measured BLS data (open circles) and
least square fit (solid line) of a
Fe(14\,nm)/Cr(0.9\,nm)/Fe(10\,nm)/Au(6\,nm)/Fe(2\,nm) sample.}
\label{nu_vs_B}
\end{figure}
Figure \ref{nu_vs_B} shows the experimental spin wave frequencies (open
circles) as a function of the external field applied
parallel to the easy [001] (bottom graph) and hard [011] direction
(top graph) of the cubic magnetocrystalline anisotropy.  Corresponding
spectra are shown in Fig.\ \ref{spectra2} below.  Only the three
dipolar modes, which are lowest in frequency, are shown.  These modes
can be classified as one acoustic and two optical modes. This classification, however,
is strictly valid only for the saturated sample.  The acoustic mode is
characterized by an in-phase precession of the magnetization in all
layers, while for the optical modes the precession of at least two neighboring
layer magnetizations is out-of-phase.  The graphs can be divided
into three distinct field regions bounded by switching events: For low field values ($B _{ext}<
35$\,mT) there is a distinct asymmetry between the Stokes and the
anti-Stokes side.  This unique feature proves an antiparallel
alignment of the magnetic moments of the two bottommost Fe layers
collinear with the external field applied in easy axis
direction.\cite{PRL_57_2442} At higher fields ($B _{ext}> 35$\,mT) the
antiferromagnetically coupled bottom Fe/Cr/Fe double layer switches
into a canted configuration with a relative angle between the layers
magnetizations of about $90^{\circ}$.  This so-called spin-flop
transition is a first order transition and can be distinguished by a
discontinuity (jump) of the mode frequencies.  The sample saturates at
an external field of about 80 and 180\,mT in the easy and hard axis
configuration, respectively.  As the coupling and anisotropy energies
are of comparable magnitude, magnetic saturation is a first order transition marked by jumps
in the frequencies for the easy axis configuration.  On the other
hand, in the hard axis configuration the saturation is approached continuously, i.e., following a
second order transition.  In this case no jumps in the mode frequencies can
be observed.  The saturation point is only marked by a minimum in the
lowest mode frequency, which is the optical mode of the
antiferromagnetically coupled Fe/Cr/Fe double layer, and a weak change
in slope of the other modes.

The hard and easy axis data have been fitted simultaneously in order to
extract the magnetic parameters of the Fe layers and the interlayer
coupling.  In order to reduce the number of adjustable parameters the
intralayer exchange stiffness and gyromagnetic ratio have been fixed
to their literature values of $D = 2.4 \times 10 ^{-17}$\,Tm$^2$ and
$\gamma / 2 \pi = 29.4$\,GHz, respectively.  The bottom (index $b$)
and central (index $c$) layers have been assumed to have equal
saturation magnetizations $M _{S(b,c)}$ and cubic anisotropies $K
_{C(b,c)}$, and the perpendicular interface anisotropies have been
assumed to be the same for the three chemically similar Fe/Ag and
Fe/Au interfaces, $K _{S (Au,Ag)}$.  $K_{S (Cr)}$ is the interface
anisotropy of the two Fe/Cr interfaces and $K_{S (ZnS)}$ of the
topmost Fe/ZnS interface.

From the very low switching field of the top layer (index $t$) of less
than 1\,mT when the field direction is reversed (see in Fig.\
\ref{nu_vs_B_lowfield} below), we can conclude that the interlayer
exchange coupling across the 6\,nm-thick Ag interlayer is negligibly
small.  Therefore, it has been set to zero.  In order to properly
describe the modes a division of all three ferromagnetic layers in
$1$\,nm-thick sheets has been employed. This is sufficient to take care of the partial nonuniformity of the modes in the $y$-direction
and of the twist of the static magnetization, which is almost negligible
for the easy axis configuration, yet has some influence on the hard
axis configuration.  The parameters extracted from
the fit are: $M _{S(b,c)} = 1.76$\,MA/m, $M _{S(t)} = 1.58$\,MA/m, $K
_{C(b,c)} = 57$\,kJ/m$^3$, $K _{C(t)} = 33$\,kJ/m$^3$, $K _{S (Au,Ag)}
= 0.5$\,mJ/m$^2$, $K_{S (Cr)} = 0.0$\,mJ/m$^2$, $K_{S (ZnS)} =
0.3\,$mJ/m$^2$, and a strength of the bilinear and biquadratic
coupling contributions across Cr of $J _1 = -0.97$\,mJ/m$^2$ and $J _2 =
-0.10$\,mJ/m$^2$, respectively.  The calculated field dependences using
these parameters are plotted as solid lines in Fig.\ \ref{nu_vs_B}.
As can be seen, the results of the calculation are in excellent overall agreement with
the experimental data.  Only the highest frequency mode differs
somehow in the canted state for the easy axis configuration.  This
small discrepancy could be due to either the too simplified $J _1$-$J
_2$ model used to describe the coupling. It has been shown to be
possibly insufficient to describe the coupling across Cr spacers,
\cite{kholin}. Another reason could be the necessary reduction of the huge number of fitting
parameters.  We also point out the notable asymmetry between Stokes
and anti-Stokes frequencies in the saturated state, especially of the
lowest optic modes in the hard axis configuration, which could not be
fully reproduced by the fit.  It is known that such an asymmetry can stem
from asymmetric interface anisotropies,\cite{PRB_41_530}, which could
not be properly taken into account in the calculation as the large
number of at least six interface anisotropy parameters (one for each
interface) is very cumbersome to fit.

\begin{figure}[hbt]
\centering\leavevmode
\includegraphics[width=0.9\linewidth,clip]{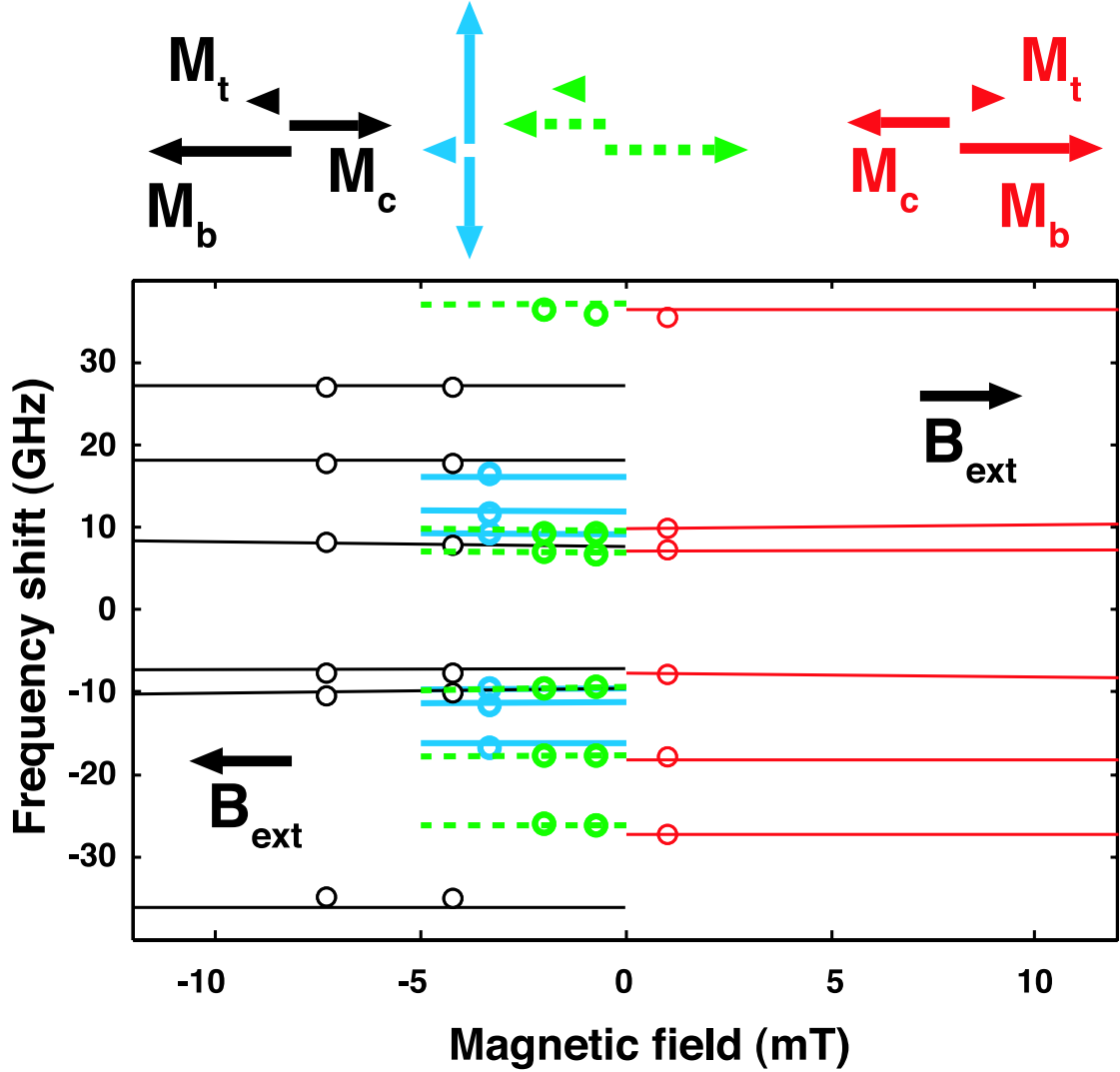}
\caption{Field dependence of measured (open circles)
and calculated (solid and dashed lines) frequencies at low fields.
The field is swept form positive to negative values. The colors of
the calculated curves correspond to the magnetization alignments
shown in the upper part.}
\label{nu_vs_B_lowfield}
\end{figure}
While an applied external field will tend to align the thickest bottom
layer and the thinnest uncoupled top layer in the field direction and,
thus, fix the magnetization directions of all layers in the
antiferromagnetic state in easy axis configuration, the situation gets
much more complex when the field approaches the zero value.  In zero external
field four directions of the magnetizations of the coupled bottom
double layer and four directions of the magnetizations of the uncoupled
top layer, each parallel to an easy axis, are possible.  However, only
nine out of these 16 possible zero field ground states can be
distinguished by the theory and the experimental setup used here, as a
reversal of the sign of the $x$-components of the static magnetization
of either the coupled double layer or the free top layer,
\textit{i.e.} transversal to the external field direction, has neither
an effect on the frequencies nor the BLS intensities due to linear
magneto-optic coupling.  The reason is that as a result only the sign
of the $z$-components of the precessional amplitudes changes.
However, they mediate neither dynamic dipolar coupling as can be concluded
form Eq.\ (A8) in Ref.\ \onlinecite{PRB_67_184404}, nor a magneto-optical
response as $m _z$ does not enter Eq.\ (\ref{phenBLS2}).  The
intensities only depend on $m _z$ via the second order magneto-optic
coupling, which has been neglected in Eq.\ (\ref{phenBLS2}), as it has only
a small impact on the intensities for our setup with a large angle of
incidence.

\begin{figure}[t]
\centering\leavevmode
\includegraphics[width=0.5\linewidth,clip]{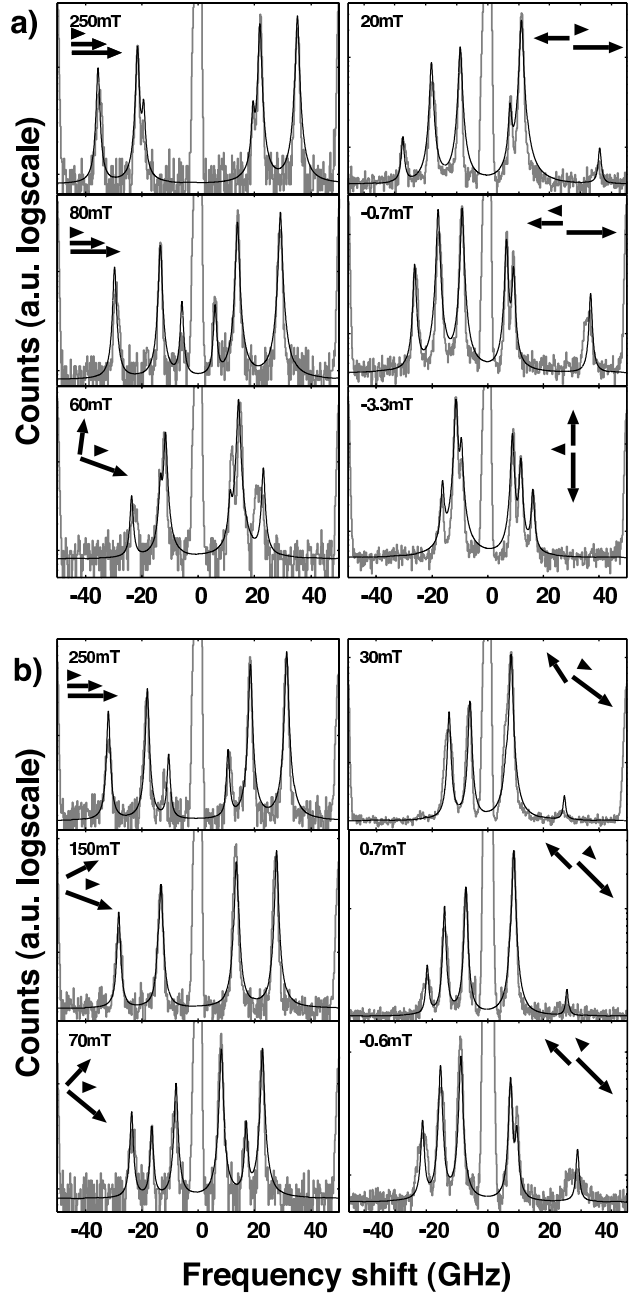}
\caption{Experimental (thick grey lines) and calculated (thin black
lines) BLS spectra at different fields applied along the easy (a) and
hard (b) axis of the magnetocrystalline anisotropy.  The computed
directions of the magnetic moments of the layers are indicated with
arrows similar to those in Fig.\ \ref{nu_vs_B_lowfield}.}
\label{spectra2}
\end{figure}
In Fig.\ \ref{nu_vs_B_lowfield} we show the experimental (open
circles) and calculated frequencies (solid and dashed lines) at low
field ($B _{ext} < 10$\,mT). To take into account the switching events, the field has been scanned from positive to
negative values and applied parallel to an easy direction.  
As the Zeeman energy has a negligible influence on
the spin waves at such low field values, the differences in frequency must
be mainly due to different alignments of the static magnetization vectors.  Four
different ground states with corresponding transitions at about -0.3,
-3, and -3.5\,mT can be observed.  By comparing the calculated and
experimental frequencies the following remagnetization behavior can be
derived: At zero field the thickest, bottommost layer points in the
positive field direction and the other layers are antiparallel to
their nearest neighbor (red state).  At a small reversed field of about
$-0.3$~mT the thin top layer switches into the field direction (green
state).  Then at about $-3$~mT the magnetizations of the bottom
double layer, initially aligned with its net magnetic moment
($M_{b}-M_{c}$) along the field direction, switches first into
the direction perpendicular to the external field (blue state) and
then completely reverses at about $-3.5$~mT (black state).  The about an order of
magnitude larger switching field of the coupled double layer as
compared to the single layer can be explained by a competition between
the pinning energy, which is proportional to the thickness, and the
Zeeman energy, which is proportional to the net magnetic moment.

While such a remagnetization sequence can be easily examined also with
integrating static methods as, for instance, MOKE or SQUID
magnetometry, the analysis of the peak areas and widths in the BLS spectra also allows one to
determine whether the sample is in a single or multiple domain state.
A multiple domain configuration can be recognized by a line broadening
and altered intensities for small domains mediating two-magnon
scattering or multiple peaks for large enough domains.  Experimental
spectra for the easy and hard axis configuration are shown as grey
curves in Fig.\ \ref{spectra2}.  Calculated spectra based on the
parameters extracted from the fit are plotted as thin black lines.
For the intensity calculations we have only used the magnetic
parameters extracted from the fit in Fig.\ \ref{nu_vs_B}, and
literature values \cite{CRC_HANDBOOK} for the indices of refraction of
the layers, intrapolated to our laser wavelength: $n_{Ag} = 0.248 +
i3.392$, $n_{Fe} = 2.595 + i3.322$, and $n_{Cr} = 2.833 + i4.450$.
The ZnS capping layer, which has only a minor influence on
the spectra has been neglected.  As the experimental FWHMs of all peaks
have approximately the same value of about 1\,GHz, which is the resolution
of the spectrometer, we have assumed a Lorentzian lineshape with this linewidth
for the calculation of the spectra.  The background level and the
absolute intensity have been adjusted manually in order to match the
experimental spectra.  The surprisingly good overall agreement of the
calculated intensities may be ascribed to the following facts: (i) the indices of
refraction of the employed materials are rather well known, (ii) the
optic properties of the capping layer, which are always somehow
uncertain due to interaction with the ambient air, \textit{e.g.}
oxidation, moisture \textit{etc.} have little influence on the spectra,
and (iii) the less well known magneto-optic coupling parameters play a negligible
role as long as they are the same for all Fe
layers.  On the other hand, the good agreement of both the frequencies
(Fig.\ \ref{nu_vs_B_lowfield}) as well as the entire spectra (Fig.\
\ref{spectra2}) proves that the theory well describes the spin wave
properties.  Hence, the calculated mode profiles are close to the actual situation in the sample,
and the sample is indeed in a single domain ground state close to the
computed one.
\section{Conclusions}
\label{conclusions}
%

We have developed a formalism for the quantitative interpretation of BLS intensities. A specific
emphasis is laid on multilayers with non-collinear alignment of the layer
magnetizations and/or non-homogeneous layer magnetization profiles in
the direction perpendicular to the layers, for instance, canted
magnetization alignments, twisted spin-configurations, or exchange
modes.  The method combines our previously presented EUTFA approach with
standard MOKE modelling and should work properly for a total thickness of the layer stack of
up to some 100\,~nm and in the frequency range of up to some 100\,GHz
accessible to BLS. Although the algebra involved is not very
demanding, a computer program is needed to determine the BLS
frequencies, mode-profiles, and intensities.  We make the source code
freely available \cite{code} in order to encourage other
researchers to analyze the BLS intensities of their experimental
spectra.

In order to show the accuracy of the formalism we have recalculated
spectra of a single, saturated Fe layer and compare it with
previous calculations.  In the experimental part we have demonstrated
that BLS spectra recorded from multilayers with three Fe layers can be
described accurately by our formalism for a large variety
of spin-configurations.  In fact, we find excellent agreement in all
cases.
Although it is clear that the well known theory of BLS should be 
able to describe the cross sections in noncollinear configurations,
experimental evidence utilizing a well
characterized model system to our knowledge has been left undone so far.

The method described here should also be easily extendable to include
out-of-plane spin-configurations and second order magneto-optic
coupling, which have not yet been taken into account in this work.
The method is well suited to gain technologically relevant,
quantitative information about spinwave mode types, the alignment of
the magnetic moments, and the magneto-optic coupling.

\section*{Acknowledgements}
The authors would like to thank J. F. Cochran, L. Giovannini and K.
Postava for helpful discussions.

%
%

\end{document}